\documentclass[aps,prd,twocolumn,natbib,floats,floatfix,amsmath,amsfonts,amssymb,nofootinbib,longbibliography]{revtex4-1}

\usepackage[hyperindex]{hyperref}
\usepackage[utf8]{inputenc}
\usepackage{graphicx}
\usepackage{dcolumn}
\usepackage{bm}
\usepackage{eufrak}
\usepackage{multirow}
\usepackage[normalem]{ulem}
\usepackage{setspace}
\usepackage{array}

\newcommand{\p}{\partial}

\newcommand{\g}{\overline{g}}

\newcommand{\h}{\bar{h}}

\newcommand{\rs}{{\rho^*}}

\renewcommand{\a}{\approx}
\newcommand{\bs}{\boldsymbol}
\newcommand{\bx}{\bar{x}}
\newcommand{\bv}{\bar{v}}
\newcommand{\aii}{{\tilde{a}_2}}
\newcommand{\aiii}{{\tilde{a}_3 }}
\newcommand{\eff}{{\mbox{\scriptsize eff}}}

\begin{document}
\title{Palatini $f(R)$ gravity in the solar system: post-Newtonian equations of motion and complete PPN parameters}

\author{J\'unior D. Toniato} \email{junior.toniato@ufes.br}
\author{Davi C. Rodrigues} \email{davi.rodrigues@cosmo-ufes.org}
\author{Aneta Wojnar} \email{aneta.wojnar@cosmo-ufes.org}
\affiliation{Núcleo de Astrofísica e Cosmologia (Cosmo-ufes), PPGCosmo \& Departamento de Física, Universidade Federal do Espírito Santo. Av. Fernando Ferrari 514, Vit\'oria-ES, Brazil.}

\date{\today}

\begin{abstract}
We perform a post-Newtonian (PN) solar system analysis for Palatini $f(R)$ theories considering finite volume non-spherical planets and with emphasis to $f(R)$ functions that are analytical about $R=0$. First we consider the Will-Nordtvedt parametrized post-Newtonian (PPN) formalism, from which the metric is shown to depend, in general, on terms not covered by the standard PPN potentials. Hence, a full analysis of the PN equations of motion is performed. From the latter we conclude that, apart from redefinitions on the internal energy and the pressure, which cannot be constrained by solar system tests, the center-of-mass orbits are the same as in general relativity.  We discuss further the physics of these redefinitions and use an argument to extend our analytical $f(R)$ results towards some non-analytical functions.
\end{abstract}

\maketitle

%\begin{footnotesize}
%\setcounter{tocdepth}{3}
%\tableofcontents
%\end{footnotesize}

\section{Introduction}

General relativity (GR) has passed many tests at astrophysical scales and is the standard gravitational theory. Nonetheless, in spite of its success, it is  relevant to consider variations on the basic assumptions of a physical theory, opening  the way for  further developments. $f(R)$ gravity theories constitute a simple extension of GR: they change the linear dependence on the Ricci scalar $R$, at the Lagrangian level, to an arbitrary function of $R$ \cite{Sotiriou:2008rp,Capozziello:2011et}. Although simple, $f(R)$ theories introduce qualitative novelties. The first of them is the breaking of the correspondence between the metric and the Palatini formalisms of GR. The Palatini approach uses an affine connection that is {\it a priori} independent from the metric. It is well known  that, for a linear $f(R)$ function, whenever the matter sector does not depend on the affine connection, the metric and Palatini formulations coincide. However, for nonlinear $f(R)$ functions, these approaches lead to remarkably different dynamics. In any spacetime region without matter, Palatini $f(R)$ can be shown to be equivalent to GR\footnote{For clarity, in this work we consider that the cosmological constant is part of GR.}, their differences may only reside inside matter distributions.

Several astrophysical aspects of Palatini gravity were considered in the literature \cite{olmo2019stellar}, such as black holes \cite{Olmo:2011ja, Olmo:2015axa, Bazeia:2014xxa, Bejarano:2017fgz, Olmo:2011np, Olmo:2012nx}, wormholes \cite{Bambi:2015zch, Olmo:2015dba, Olmo:2015bya} and stellar objects \cite{Kainulainen:2006wz, Reijonen:2009hi, Panotopoulos:2017jdc, Pannia:2016qbj, Wojnar:2017tmy}. Although problems for stars in Palatini $f(R)$ gravity were reported \cite{Barausse:2007ys, Barausse:2007pn, Pani:2012qd, Sham:2013sya}, these issues depend on matter assumptions that need not to be true \cite{Olmo:2008pv, Mana:2015vqa, Wojnar:2018hal, Sergyeyev:2019aul}. It was also discussed in \cite{Capozziello:2011gw} that the initial value problem is well formulated when standard matter sources are present, while the well-posedness of the Cauchy problem must be considered case by case.

For solar system physics, the simplest test consists of modeling the Sun as a central mass and treat the planets as test particles. Since Palatini $f(R)$ is equivalent to GR in vacuum, the spacetime solution is Schwarzschild-de Sitter, and the test particles will follow standard geodesics (these results are reviewed in Section \ref{palatini}). Hence the parametrized post-Newtonian (PPN) approach of Eddington-Robertson-Schiff, which is based on the metric solution in vacuum, with the planets as test particles, yields the same predictions of GR. That is, in this context, no constraints on the function $f(R)$ are found, apart from possible constraints due to the cosmological constant \cite{Liang:2014vma}. However, this test is neither robust nor precise since it depends on a very particular geometrical setting. It is interesting that GR happens to yield the same orbits for both test particles and for the center of mass of finite volume planets; but this property does not hold in general for alternative theories of gravity \cite{Nordtvedt:1968qs, Will:1971zzb, will1993theory}.

A more robust and precise way to confront theory with observational data is through the Will-Nordtvedt PPN formalism \cite{will1993theory, Will:2014kxa}. It is based on perfect fluids, not on point particles; hence planets are finite-volume objects with finite density. This framework assumes an asymptotically flat space and considers a post-Newtonian (PN) metric written as an expansion on powers of $v/c$ (i.e.,  the ratio between the velocity field magnitude $v$ and the speed of light $c$). This expansion depends on gravitational potentials whose dimensionless coefficients, the PPN parameters, can be constrained by observational data. The number of gravitational potentials considered by this PPN approach is limited, this due to practical reasons. Consequently, although the formalism is sufficiently general to include GR and a variety of modified (or extended) gravity theories (e.g., some scalar-tensor theories, bigravity models \cite{Will:2014kxa}, a class of renormalization group-based extensions of GR \cite{Toniato:2017wmk}...), there are yet several examples of theories that are not covered by the formalism, such as Horndeski \cite{Hohmann:2015kra} and general $f(R)$ theories \cite{Clifton:2008jq,Olmo:2005hc}. When facing a theory with potentials not covered by the standard PPN approach, one cannot infer physical bounds from the values of the standard PPN parameters that appear in the theory, since the additional potentials may influence the values of the standard PPN parameters. Therefore, if additional potentials appear, it is safer to consider physical bounds directly based on the PN field equations. The latter is the approach that we use here. We are not aware of other works on Palatini $f(R)$ gravity that evaluate physical constraints directly from the field equations.

Instead of trying to uncover the PPN bounds for an arbitrary $f(R)$ theory, we concentrate on precise statements that can be drawn from a relevant class of functions, which is composed by $f(R)$ functions that are analytical about $R=0$ (i.e., those that can be Taylor expanded about zero). Therefore, models that include terms like $1/R$, which have attracted considerable attention in the past due to its possible dark energy implications (see \cite{Olmo:2011uz} for a review), are not our main concern here. Analytical $f(R)$ functions lead to metric expressions that are closer to the standard Will-Nordtvedt PPN formalism, which allow us to do a detailed PN analysis that was not done before. Also, from that analysis, we further detail here an interpretation of some Palatini terms that appear in the metric. Curiously, our detailed PN analysis, together with its corresponding physical interpretation, does not lead to stronger bounds, but to the absence of bounds coming from the solar system analysis. We also consider laboratory fluid tests, which were commented in \cite{Olmo:2005hc} as a Palatini $f(R)$ gravity test, but we find that the constraints from such tests depends on the physical interpretation of some terms, and we argue in favor of an interpretation that leads to no such bounds.

\section{Palatini $f(R)$ gravity review and point-particle planets}\label{palatini}
The $f(R)$ gravitational theories using Palatini approach considers spacetime metric $g_{\mu\nu}$ and affine connection ${\Gamma}^{\lambda}_{\mu\nu}$ as independent objects of a manifold, which we consider torsionless (i.e., ${\Gamma}^{\lambda}_{\mu\nu} = {\Gamma}^{\lambda}_{\nu\mu}$)\footnote{See \cite{Olmo:2011uz} for a review on the case with torsion.}, which is the most studied case. For a given vector field $A^\sigma$, the Riemann  tensor $R^\lambda_{\mu \sigma \nu}$ is defined by $R^\lambda_{ \sigma \mu \nu} A^\sigma= [ \nabla_\mu, \nabla_\nu] A^\lambda$, where $\nabla_\mu$ is the covariant derivative based on the affine connection ${\Gamma}^{\lambda}_{\mu\nu}$. The Ricci tensor and Ricci scalar are respectively given by $R_{\mu \nu} \equiv R^\lambda_{  \mu \lambda \nu} $ and $R \equiv g^{\mu \nu} R_{\mu  \nu}$. Since $g_{\mu \nu}$ and ${\Gamma}^{\lambda}_{\mu\nu}$ are not assumed to be compatible, contrary to Riemannian geometry, $\nabla_\mu g_{\nu \lambda} \not=0$. The action for Palatini $f(R)$ gravity with matter is
\begin{equation}
S[g,\Gamma, \Psi] = \frac{1}{2\kappa}\int f(R) \sqrt{-g} \, d^4x + S_{\mbox{\tiny matter}}[g,\Psi], \label{action}
\end{equation}
where $\kappa$ is some coupling constant, $\Psi$ represents an arbitrary set of fields that describe any other interactions apart from gravity. One could generalize the matter action $S_{\mbox{\tiny matter}}$ above to include a dependence on $\Gamma$. Such extension is expected to be relevant in the context of fermionic interactions, not in the context of large scale classical fluids, hence we will not consider it.

The action variations with respect to $g$ and $\Gamma$ lead respectively to 
\begin{align}
	&f'(R)R_{\mu\nu}-\dfrac{1}{2}\,f(R)g_{\mu\nu}=\kappa\, T_{\mu\nu}\,,\label{feq}\\[.1in]
	&\nabla_\lambda\left(\sqrt{-g}f'(R)g^{\mu\nu}\right)=0\,.\label{cov}
\end{align}

Similarly to GR, the action variation with respect to $\Psi$, together with diffeomorphism invariance of $S_{\mbox{\tiny matter}}$, implies that
\begin{equation}
	\nabla_{\!C}^\mu T_{\mu \nu} =0 \, , \label{consv}
\end{equation}
where 
\begin{equation}
	C_{\mu \nu}^\lambda = \frac 12 g^{\lambda \sigma}\left(g_{\mu \sigma, \nu} + g_{\sigma \nu, \mu} - g_{\mu \nu, \sigma}\right)
\end{equation}
is the Christoffel symbol. The commas indicate partial derivatives. This implies that test particles will follow geodesics determined from the connection $C$, not the fundamental $\Gamma$ connection. 

A stronger similarity between Palatini $f(R)$ gravity and GR comes from the trace of eq.~\eqref{feq},
\begin{equation}
	f'(R) R - 2 f(R) = \kappa T \,, \label{tracefeq}
\end{equation}
which shows that $R$ can be algebraically expressed as a function of $T$. This is in  contrast to the metric $f(R)$ case, in which the relation between $R$ and $T$ comes from a differential equation, implying an additional degree of freedom. In general, eq.~\eqref{tracefeq} will have more than one $R$ solution for a given $T$. For $T_{\mu \nu} =0$, one finds $T=0$ and $R$ must be a constant. Let $R_{s}$  denote one of the Ricci scalar solutions, indexed by $s$, in a spacetime region with $T_{\mu \nu}=0$. We consider that $f'(R_s) \not=0$, otherwise there would be no relation between the metric and the connection at the field equations level \cite{Allemandi:2005tg}. Hence, from  eq.~\eqref{tracefeq}, 
\begin{equation}
	f'(R_s) R_s -  2 f(R_s)=0	 \label{traceVac}
\end{equation}
and the field equation \eqref{feq} becomes
\begin{equation}
	R_{\mu \nu}(\Gamma) = \frac 14 g_{\mu \nu} R_s \, . \label{RGammavac}
\end{equation}
The dependence on $\Gamma$ was written to recall that the above is a differential equation for $\Gamma^\lambda_{\mu \nu}$, not for the metric. From eq.~\eqref{cov}, one can write the above as a differential equation for a metric $\overline g_{\mu \nu}$ that satisfies $\nabla_\lambda \overline g_{\mu \nu} = 0$, namely
\begin{equation}\label{gbar}
\g_{\mu\nu}=f'(R)\,g_{\mu\nu} \, .
\end{equation}
This implies that $\Gamma^\lambda_{\mu \nu} $ can be interpreted as the Christoffel symbol of the metric $\overline g_{\mu \nu}$,
\begin{equation}
 	\Gamma_{\mu \nu}^\lambda = \frac 12 \overline g^{\lambda \sigma}\left(\overline  g_{\mu \sigma, \nu} + \overline  g_{\sigma \nu, \mu} - \overline  g_{\mu \nu, \sigma}\right) \, .
\end{equation}
Hence eq.~\eqref{RGammavac} can be written as differential equation for $\overline g_{\mu \nu}$, as follows (see also  \cite{Allemandi:2005tg}),
\begin{equation}
	R_{\mu \nu}(\overline g) = \frac 14 \, \overline g_{\mu \nu} \frac {R_s}{ f'(R_s)} \, . \label{RGammavac1}
\end{equation}

For  GR in vacuum and using $\overline g_{\mu \nu}$ as the metric, the Einstein field equation can be written as $R_{\mu \nu}(\overline g) = \overline g_{\mu \nu} \Lambda $. Therefore, by identifying an effective cosmological constant as
\begin{equation}
	\Lambda_\eff =  \frac 14 \frac{R_s}{f'(R_s)} =\frac{f(R_s)}{2 f'^2(R_s)}\, , \label{Leff}
\end{equation}
one finds that, in vacuum, Palatini $f(R)$ gravity and GR have the same field equations and the same geodesic equations.\footnote{For GR the geodesics are found from the Christoffel symbol associated to the same metric that appears in the field equations ($\overline g_{\mu \nu}$), denoted by $\Gamma^\lambda_{\mu \nu}(\overline g)$. For Palatini, the connection to be used is the Christoffel symbol of a conformally related metric, $C^\lambda_{\mu \nu}(g)$. Since,  in vacuum, $\bar g_{\mu \nu}$ and $g_{\mu \nu}$ only differ by a constant conformal rescaling, these connections coincide  $\Gamma^\lambda_{\mu \nu} = C^\lambda_{\mu \nu}$.} One also notes that,  in order to find $\Lambda_\eff =0$, it is sufficient and necessary that $R_s=0$ or, equivalently, that $f(R_s)=0$.

The above conclusions imply that, in case the Sun is assumed to be spherically symmetric and the planets are considered test particles, the metric solution outside the Sun is a Schwarzschild-de Sitter solution. In this case, the single non-trivial constraint possible to be applied to Palatini refers to a bound on the value of $\Lambda_\eff$, and consequently on $R_s/(2 f'(R_s))$. Such bound only  implies that  $\Lambda_\eff \lesssim 10^{-43} \mbox{ m}^{-2}$ \cite{Liang:2014vma}. We remark that this bound cannot be found from Will-Nordvedt PPN formalism, since the latter assumes that the space is asymptotically flat. 

In this context of point particles, and considering that $\Lambda_\eff$ is dynamically irrelevant, the Eddington-Robertson-Schiff parameters $\gamma_{\mbox{\tiny ERS}}$ and $\beta_{\mbox{\tiny ERS}}$ are both equal to 1, just like GR.

\section{Post-Newtonian Expansion and potentials} \label{sec:ppn}
As nicely summarized in Ref.~\cite{PoissonWill}, ``Post-Newtonian theory is the theory of weak-field gravity within the near zone, and of the slowly moving systems that generate it and respond to it''. It is suitable to describe solar system dynamics and compare theoretical predictions with experiments and observational data. To implement this framework to Palatini $f(R)$, we follow Refs. \cite{will1993theory, Will:2014kxa, PoissonWill}. In this section, we first present a systematic review of the PN approximation and then apply it to Palatini gravity.

The metric is expanded about a flat spacetime,
\begin{equation}\label{metricexp}
g_{\mu\nu}=\eta_{\mu\nu}+h_{\mu\nu}\,,
\end{equation}
where $\eta_{\mu\nu}$ is the background Minkowski metric and $h_{\mu \nu}$ contains all the perturbations. The matter content is described by a perfect fluid, using $c=1$,
\begin{align}
	T^{\mu\nu}=\left(\rho+\rho\Pi+p\right)u^{\mu}u^{\nu}+pg^{\mu\nu}\,,
\end{align}
where $\rho$ is the mass density, $\Pi$ is the fluid's internal energy per unity of mass, $p$ is the pressure and $u^{\mu}=u^{0}(1,\bs v)$ is the four-velocity of the fluid. The components of the stress-energy tensor can be expanded in orders of $v$, taking into account also that $p/\rho\sim \Pi\sim O(2)$ for the typical matter in the solar system. Here and after, we use the notation $O(N)$ to represent quantities of order $v^{N}$. Time derivatives count as one order increase since the dynamical timescale in solar system is governed by the motion of planets, i.e. $\p_0\sim O(1)$. The fluid dynamics is subjected not only to eq.~\eqref{consv}, but also to the conservation of rest mass density, 
\begin{equation}\label{consmatter}
\nabla_{\mu}(\rho u^{\mu})=0 \, .
\end{equation}
This equation can be re-expressed as an effective flat-space continuity equation as follows,
\begin{equation}\label{cont}
\p_t(\rs)+\p_i(\rs v^{i})=0,
\end{equation}
with Latin indices for the spatial components and
\begin{equation}\label{rstar}
\rs\equiv {u^{0}}\sqrt{-g}\,\rho.
\end{equation}
Since $\rs$ is the conserved density, in the sense of \eqref{cont}, it is more convenient to use it to express the energy-momentum tensor components. As will become clear later, having the gravitational potentials expressed in terms of the conserved density will lead to an easy way to integrate the equations of motion of massive bodies.

For any $f(R)$ function that is analytic at $R=0$, one can write
\begin{equation}\label{poly}
f(R)=\sum_{n=0}^{\infty}a_nR^{n},
\end{equation}
where $a_0, a_1, a_2, ...$ are constants. In order to proceed with PPN formalism, we impose that space should be asymptotic flat, thus, from eq.~\eqref{Leff}, we find $a_0=0$. Since $\kappa$ is arbitrary [see eqs.~(\ref{action}, \ref{feq})], without loss of generality we can set $a_1=1$. Therefore, since $R$ is at least of order two, eq.~\eqref{tracefeq} can be written as
\begin{equation}
R= -\kappa\,T + O(6).
\end{equation}
The $a_2$ contribution cancels out from the above equation, while $a_n$ with $n \geq 3$ only contributes to $O(6)$ or higher order. It is important to stress that, from the above result, for analytic $f(R)$ functions and up to the first PN order, there are not many, but a single solution to  eq.~\eqref{traceVac}.  Moreover, since we have set $a_0=0$, the single vacuum solution for $R$ is $R =0 + O(6)$.

With the above, eq.~\eqref{feq} can be expressed as
\begin{eqnarray}\label{feq2}
R_{\mu\nu}=&& \ \kappa\left(T_{\mu\nu}-\dfrac{1}{2}\,g_{\mu\nu}\,T\right) \ +\nonumber\\
&&+ \ 2\kappa^{2}a_2 T\left(T_{\mu\nu}- \dfrac{1}{4}\,g_{\mu\nu}\,T\right) + O(6).
\end{eqnarray}
In what follows, we will solve the equation above, order by order, to obtain the PN metric of the considered class of $f(R)$ Palatini gravitational theories. Following the standard PN procedures, all geometrical quantities are at least of order two. Moreover, $h_{00}$ needs to be known up to $O(4)$, $h_{0i}$ up to $O(3)$ and $h_{ij}$ up to $O(2)$.

\subsection{Second order}
We start by considering the leading terms in eq.~\eqref{feq2}. In the right hand side, up to second order, $g_{\mu\nu}$ can be replaced by $\eta_{\mu\nu}$, $u^{0}\a 1$ due to the normalization of $u^{\mu}$ and $\rs\a\rho$ according to eq.~\eqref{rstar}. Consequently, the only component of energy-momentum tensor to be considered is $T_{00}\a -T\a \rs$ and one finds $R_{0i} \approx 0$. Therefore,
\begin{align}
	R_{00} =& 4 \pi  \rho^* +O(4)\, ,  \label{R2left00}\\[2ex]
	R_{ij}  = & 4 \pi   \rho^*  \delta_{ij} + O(4) \, . \label{R2leftij}
\end{align}
In the above, and henceforth, we set the mass dimension such that 
\begin{equation}
	\kappa = 8 \pi.		
\end{equation}

As previously introduced, the Ricci tensor is a function of the connection $\Gamma_{\mu \nu}^\lambda$, which in turn can be expressed as the Christoffel symbol of the auxiliary metric $\overline g_{\mu \nu}$. Instead of directly expressing $R_{\mu \nu}$ as an expansion on $h_{\mu \nu}$, it is convenient to first express it as an expansion on $\overline h_{\mu \nu}$, which is defined by
\begin{equation}
	\overline g_{\mu \nu} = \eta_{\mu \nu} + \overline h_{\mu \nu} \, .
\end{equation}
By using eq.~\eqref{gbar}, one can finally express all the results with respect to $h_{\mu \nu}$. We take the opportunity to note that the asymptotic flatness property of $g_{\mu \nu}$ is inherited by the auxiliary metric $\overline g_{\mu \nu}$ due to the analyticity of $f(R)$. Indeed, assuming that far away from the main system (``at infinity'') we have vacuum, then $R = 0$ and $f'(0)=1$\footnote{Apart from $O(6)$ terms, which are not relevant.}. Hence, eq.~\eqref{gbar} implies that the asymptotic limits of the metric and the auxiliary metric are identical.

The leading terms of $R_{\mu\nu}$ (which are of  second order) are
\begin{align}
	R_{00}&\a - \frac{1}{2} \nabla^{2}\h_{00} \,,  \label{R2right00} \\[2ex]
	R_{ij}&\a \frac{1}{2} \Big(- \nabla^{2}\h_{ij} + \p_{ij}\h_{00} - \p_{ij}\h^{k}_{k} + 2\p_{k(i}\h^{k}_{j)} \Big), \label{R2rightij}
\end{align}
with $A_{(ij)}\equiv(A_{ij}+A_{ji})/2$.

By combining eqs.~(\ref{R2left00}, \ref{R2leftij}, \ref{R2right00}, \ref{R2rightij}), we find
\begin{align}\label{eq002}
&\nabla^{2} \h_{00}{} = - 8 \pi \rs \,, \\[2ex]
&\nabla^{2}\h_{ij} + \p_{ij}\h^{k}_{k} - 2\p_{k(i}\h^{k}_{j)} -\p_{ij}\h_{00} = - 8 \pi \rs \delta_{ij}. \label{eqij}
\end{align}

By choosing coordinates such that off-diagonal components of $\h_{ij}$ are zero, which is a standard procedure within PPN, it is possible to write the $\h_{00}$ and $\h_{ij}$ general solutions as
\begin{align}\label{hb00}
	\h_{00} &= 2U +O(4), \\[2ex]
	\h_{ij} &= 2U\delta_{ij} +O(4). \label{hbij}
\end{align}
From eq.~\eqref{eq002}, and using that the auxiliary metric are asymptotically flat, $U$ is found to be (negative of) the Newtonian potential,
\begin{equation}
	U=  \int\dfrac{\rs(t,\bs x') \,d^{3}x'}{|\bs x-\bs x'|}.
\end{equation}

The above is sufficient to specify the auxiliary metric solution up to second order. With respect to the original metric $g_{\mu \nu}$, the above solutions read, from eq.~\eqref{gbar},
\begin{align}\label{metric2}
g_{00}\a& -1 + 2U +  2 \aii  \rs,\\[2ex]
g_{ij} \a& \ \delta_{ij}+  2 U \delta_{ij} -   2 \aii  \rs \delta_{ij},\\[2ex]
g_{0i} \a& \ 0\,,
\end{align}
where $\aii \equiv 8\pi a_2$.

\subsection{Third order}
Moving to the third order, only the $0i$-components of the field equation contribute, 
\begin{equation}\label{eq0i}
\nabla^{2}\left(\h_{0i} +4V_{i}\right) + 4\p_{ti}U - \p_{ik}\h^{k}_{0}\a 0,
\end{equation}
where $V^{i}$ is a vector potential defined by
\begin{equation}\label{vi}
V^{i}= \int\dfrac{\rs(t,\bs x')v^{i}}{|\bs x-\bs x'|}\,d^{3}x',\quad \nabla^{2}V^{i}=-4\pi\rs\, \bar{v}^{i}.
\end{equation}
To solve this equation, we impose another standard  gauge condition for PPN analyses, which is given by
\begin{equation}\label{gauge}
\p_{k}\h_{0}^{k}=2\p_{t}\h_{00}.
\end{equation}
The solution is then easily obtained to be
\begin{equation}\label{hb0i}
\h_{0i}\a-4V^{i},
\end{equation}
and, through \eqref{gbar}, the correspondent physical metric is,
\begin{equation}\label{metric3}
g_{0i}=-4V^{i} + O(5).
\end{equation}

The gauge choice \eqref{gauge} is used in the more modern approach of Ref.~\cite{PoissonWill}, but not in the classical PPN formalism \cite{will1993theory}. However, as it will become clear later, Palatini $f(R)$ does not fit with the parametrized formalism and the gauge adopted here is more suitable for derivation of equations of motion. 

\subsection{Fourth order}

Since the single metric component that needs to be known up to the fourth order is  $g_{00}$, we consider, 
\begin{align}
R_{00}\a& - \frac{1}{2} \nabla^{2}\h_{00} + \frac{1}{4} \Big(-2\p_{tt}\h^{i}_{i} + 4\p_{0i} \h_{0}^{i} - |\vec{\nabla} \h_{00}|^{2} \ - \nonumber\\[2ex]
&  - \ \p_{i}\h^{k}_{k}\p^{i}\h_{00} + 2\p^{i}\h_{00} \p_{k}\h^{k}_{i} + 2 \h^{ik}\p_{ik}\h_{00}\Big).
\end{align}
For the matter part, we use expressions \eqref{metric2} and \eqref{metric3} and compute $u^{0}$ and $\rs$ up to the second order, namely
\begin{align}
	u^{0}&= 1 + \left(2U + v^{2} +\aii\rs\right) +O(4),\\[2ex]
	\rs&= \rho\left(1+3U +\dfrac{v^{2}}{2}-3\aii\rs\right) +O(4).
\end{align}
With the above, we find
\begin{eqnarray}
\nabla^{2}\h_{00}= & -8\pi\rs\left(1 +\dfrac{3}{2}v^{2} -U +\Pi +3p -2\aii\rs \right) -\nonumber \\ & -2\p_{tt}U.
\end{eqnarray}
By using that the auxiliary metric is asymptotically flat, the general solution reads
\begin{equation}\label{hb004}
\h_{00}=-2U +2 \left (\psi +\dfrac{1}{2}\p_{tt}\chi -U^{2}-2\aii \Phi_P \right),
\end{equation}
with 
\begin{equation}\label{psi}
   \psi\equiv\dfrac{3}{2}\,\Phi_1-\Phi_2+\Phi_3+3\Phi_4,\\[1ex]
\end{equation}
and the PN  potentials that appear above are:
\begin{align}
   \Phi_1=\int\dfrac{\rs{}' v'^{2}}{|\bs x- \bs x'|}\,d^{3}x',\quad \Phi_2=\int\dfrac{\rs{}' U'}{|\bs x- \bs x'|}\,d^{3}x',\\[1ex]
   \Phi_3=\int\dfrac{\rs{}' \Pi'}{|\bs x- \bs x'|}\,d^{3}x',\quad \Phi_4=\int\dfrac{p'}{|\bs x- \bs x'|}\,d^{3}x' \, ,\\[1ex]
   \chi=\int\rs{}'|\bs x- \bs x'|\,d^{3}x',\quad \Phi_{P}=\int\dfrac{\rs{}'^{2}}{|\bs x- \bs x'|}\,d^{3}x' \, .
\end{align}
The potential $\Phi_{P}$, which we call Palatini potential, is the single one which is absent from GR  (it is the only one that appears multiplied by $\aii$). However, this is not the final answer, since the result above only concerns the auxiliary metric $\g_{\mu \nu}$. To find the corresponding expansion for  the metric $g_{\mu \nu}$, we use eq.~\eqref{gbar}, implying\footnote{We note that the potential $\chi$ does not appear in the approach of Refs.~\cite{will1993theory, Will:2014kxa}, but it appears in Ref.~\cite{PoissonWill}, due to a gauge difference.}
\begin{align}
	g_{00} \a & -1 + 2 U + 2\psi +\p_{tt}\chi  -  2U^2 + \aii \big( 2  \rho^*- 4 \Phi_{P}    - \notag \\ 
&  - 10  U\rho^* +  v^2 \rho^* - 2  \Pi  \rho^* + 6   p  -  2 \aii {\rho^*}^2 \big)  + 3 \aiii  {\rho^*}^2, \notag
\end{align}
where we have defined $\aiii=64\pi^{2}a_{3}$. 

Considering the above expression for $g_{00}$ there are a couple of remarks to be done: i) at the Newtonian order, there is a Palatini correction term that is added to the Newtonian potential and is given by $ \aii \rho^*$, this term is in general relevant for the internal stability of a given body, say a planet, star or galaxy, but it does not change the body center of mass trajectory (this is further detailed in Section \ref{sec:eqofmot}). ii) Although $\g_{\mu \nu}$ does not depend on $a_3$, this term  has, in general, a dynamical effect at the first PN order; while $a_n$ with $n\geq 4$ has no dynamical impact up to the first PN order. This implies that these results can be trivially generalized to a non-analytical $f(R)$ only in case the non-analytical part of $f(R)$ can be shown to be of higher order than $R^3$. iii) If there is no matter ($\rho^*=p=0$) in a given spacetime region, then in that region Palatini gravity should coincide with GR, as we have previously shown. Nonetheless, there is a single term in $g_{00}$ that does not appear in GR and does not vanish if $\rho^* \rightarrow 0$ and $p \rightarrow 0$ locally: the $\Phi_P$ potential. Therefore, there must exist a redefinition such that this term is absorbed by the other ones.  Indeed, if the microphysics of the gravitational source is not sufficiently well constrained, while the orbits are assumed to be well constrained, one can redefine $\Pi$ and $p$ as $\Pi_\eff$ and $p_\eff$, such that
\begin{equation}
	\rho^* \Pi_\eff + 3 p_\eff = \rho^* \Pi + 3 p - 2 \aii {\rho^*}^2 \, . \label{PipEff}
\end{equation} 
This redefinition implies that Palatini $f(R)$ gravity in vacuum can be recast as GR in vacuum with a perfect fluid as the gravitational source, but with a different perfect fluid description. This observation does not hold inside matter.\footnote{Rastall gravity \cite{Rastall:1973nw} is an example of a theory that was proposed as a modified gravity theory, but it can be fully recast as a modified perfect fluid \cite{Visser:2017gpz}. Apart from vacuum, this is not the case of Palatini $f(R)$.} It was  only introduced to explicitly recover GR in vacuum, eliminating the $\Phi_P$ potential from the metric.\footnote{One can also directly check that this same redefinition also removes $\Phi_P$ from the geodesics up to the 1PN order, eq.~\eqref{1PNEuler}.} This redefinition will be further discussed in the context of  conserved quantities, in Sec.~\ref{sec:conserved}.

Collecting together the previous results,
\begin{align}
g_{00} \a & -1 + 2 U + 2\psi +\p_{tt}\chi  -  2U^2 - 4\aii \Phi_{P} + 3 \aiii  {\rho^*}^2   \notag \\
	& + \aii \rs\!\left( 2 - 10U +  v^2\!  - 2 \Pi  + 6p/\!\rs \! -  2 \aii \rho^* \right)\!, \\[1ex]
	g_{0i} \a& -4V^{i},\\[1ex]
	g_{ij} \a & \delta_{ij}+2\left(U -\aii\rs \right)\delta_{ij}.
\end{align}
Besides the $\Phi_p$ potential, the metric above contains various terms that are outside the standard Will-Nordvedt parametrization; implying that it is not correct to infer any immediate bound on any parameter.

As expected, when $\aii$ and $\aiii$ go to zero, the 1PN metric of GR is recovered. Also, we note that our result for the metric is not in conflict with previous analysis on the PN limit of $f(R)$ Palatini theories using the scalar-tensor equivalence \cite{Olmo:2005hc,Olmo:2005zr,Sotiriou:2005xe}, when the latter are particularized to an analytic $f(R)$ function. On the other hand, the relevant particular case here studied allow us to  present all the terms in explicit form and to proceed rigorously with the PPN formalism.  

Before considering further details inside matter, we comment on the trajectory of electromagnetic waves (or light in particular).  All the experiments used to constrain the trajectories and time delays of electromagnetic waves do not depend on a given medium, they only depend on the vacuum spacetime. Hence, the only metric that such experiments can probe for Palatini $f(R)$ gravity is the Schwarzschild-de Sitter metric. Such electromagnetic experiments cannot infer any difference with respect to GR. It should be recalled the following  natural assumptions: i) $\Lambda_\eff$ is assumed to be sufficiently small such that asymptotic flatness is in agreement with the solar system; ii) there is no data and theory available capable of deriving the gravitational mass, up to the 1PN order, of the Sun or the planets from the microphysics of their internal structure, this independently from gravitational experiments. Therefore,   the $\gamma$ value within Palatini $f(R)$ gravity is precisely 1, just like GR. We remark that this result on $\gamma$ does not depend on $f(R)$ being analytical. This is in contrast with a result from Ref.~\cite{Olmo:2005zr}: the reason for the difference comes from non-asymptotically flat effects which should not be considered in PPN, unless a proper explanation on the validity of the method is included.\footnote{Indeed, the meaning of light bending in the presence of a cosmological constant has several subtleties and it has generated a debate in the literature  (see e.g., \cite{Rindler:2007zz, Biressa:2011vy, Piattella:2015xga, Faraoni:2016wae}). The Will-Nordtvedt PPN formalism was not built to handle such issues nor non-asymptotically flat spacetimes in general.}

To determine the equations of motion of massive, finite-volume bodies, further details are necessary. The first step is to examine the PN hydrodynamics in order to find the conserved quantities.

\section{Conserved quantities}\label{sec:conserved}
The re-scaled density $\rho^*$ allows one to define the  fluid material mass in a given volume as follows,\footnote{We use the term ``material mass'' in the same sense used in Ref.~\cite{PoissonWill}. It is different from inertial mass, passive gravitational mass and active gravitational mass. }
\begin{equation}\label{mass}
m=\int\rs d^{3}x \, .
\end{equation}
Hence, for an isolated body, by using the continuity equation for $\rho^*$ \eqref{cont}, mass is a constant,
\begin{equation}\label{mcons}
\dfrac{dm}{dt}=0\, .
\end{equation} 
In the above, it was used that the density $\rho^*$ is zero at the boundary of the integration volume. The statement above only depends on the fluid definition, not on the used gravity model. The situation is different for energy conservation.

From the energy-momentum tensor conservation \eqref{consv}, we write
\begin{equation}\label{divt}
\p_\nu(\sqrt{-g}T^{\mu\nu})+ \Gamma^{\mu}_{\lambda\nu}(\sqrt{-g}T^{\lambda\nu})=0.
\end{equation}
For $\mu=0$ and up to $O(1)$, the mass continuity equation \eqref{cont} is recovered; but considering this equation at the third order $O(3)$, one gets
\begin{align}\label{localenergy}
\rho^{*}\frac{d}{dt}\left(\frac{v^{2}}{2}+ \Pi \right) &+  \partial_{i}(pv^{i}) \ - \nonumber\\ 
&- \ v^{i} \rho^{*}{} \partial_{i}U  -  \frac{\aii}{2} v^{i} \partial_{i}(\rho^{*2}) = 0\, .
\end{align}
This result can be expressed as an energy conservation statement for an isolated body,
\begin{equation}\label{econs}
\frac{dE}{dt}=0,
\end{equation}
with
\begin{equation}\label{energy}
E\equiv\int\left(\frac{1}{2}\rs v^{2}  + \rs\Pi - \frac{1}{2}\rs U -\frac{\aii}{2}\rs{}^{2}\right)d^{3}x.
\end{equation}

To show the above result, we used that for an  arbitrary function $f(t,x)$, and considering the integration on a volume that only includes an isolated body (i.e., $\rho^*=0$ at the boundary), we have
\begin{align}\label{dt}
\frac{d}{dt}\int\rs f\,d^{3}x &= \int \left( \dot {\rho^*}  f  + \rho^* \dot f \right) d^3x \notag \\[2ex]
&= \int\rs\frac{df}{dt}d^{3}x \, .
\end{align}
The continuity equation for $\rho^*$, eq.~\eqref{cont},  was used above. This is sufficient for explaining the two first terms in eq.~\eqref{energy}. The third term, which depends on $U$, is derived in detail in Ref.~\cite{PoissonWill}. For the fourth term, which is the single additional contribution from Palatini $f(R)$, it is used that
\begin{equation}
	\int v^i \partial_i ({\rho^*}^2) \, d^3x = 2 \int \rho^* \dot \rho^* \,  d^3x = \frac{d}{dt} \int \rs^2 \, d^3 x \, .
\end{equation}

The total mass-energy of the fluid is then defined as,
\begin{equation}
M=m+E,
\end{equation}
and, through Eqs. \eqref{mcons} and \eqref{econs}, it satisfies $dM/dt=0$. 

Considering the case in which Palatini gravity in vacuum yields the same gravitational mass of GR, the energy $E$ from Palatini can be expressed as energy from GR derived from a body whose  effective internal energy is given by
\begin{equation}
	\Pi_\eff = \Pi - \frac{\aii}{2} \rs \, . \label{PiEff}
\end{equation}
Together with eq.~\eqref{PipEff}, this implies that
\begin{equation}
	p_\eff = p - \frac{\aii}{2} \rs^2 \, . \label{pEff}
\end{equation}

Considering now the case $\mu = i$ in eq.~\eqref{divt}, at the first nontrivial order one finds
\begin{align}
	\rs\frac{dv^{j}}{dt} &=\rs\p^{j}U-\p^{j} \!\left(p - \frac{\aii}{2} \rs^2 \right)  \notag\\[2ex]
	&= \rs\p^{j}U-\p^{j} p_\eff  \, . \label{neuller}  
\end{align}
Although at the first sight the $\aii$ term above could be seen as a correction to the Euler equation that would depend ultimately on how the pressure $p$ is defined. For many physical experiments with fluids, there is no compelling reason to consider $p$ to be the ``true pressure''. In the above, we divided the Euler equation into two parts: the first depends on $U$ and is the only term capable of generating a force in vacuum. The second term depends on the local values of $p$ and $\rho^*$, hence it is non-null only inside the fluid and it can be interpreted as an effective pressure. This effective pressure interpretation is further supported by eq.~\eqref{pEff}. For the majority of the Euler equation applications, one does not define the pressure $p$ from the knowledge of the fluid microphysics: the pressure is commonly taken to be a force per unity of area that is locally generated by the fluid; hence, eq.~\eqref{neuller} is not necessarily an extended Euler equation, it is the Euler equation. In this interpretation,  $p_\eff$ is the physically relevant pressure while $p$ just denotes a component of the energy-momentum tensor without direct physical meaning. We also stress that in practice what determines pressure is not its place in the energy-momentum tensor, but its place in the matter field equations, which are found from eq.~\eqref{divt}. This is not the single possible interpretation for $p_\eff$ and $p$, but, in order to consider eq.~\eqref{neuller} as a modified Euler equation, 
the first proper physical meaning for $p$ should be provided (and this step is commonly skipped). Hence, in particular, unless further information on how to macroscopically measure $p$ independently from the Euler equation is provided, we disagree with the statement of Ref.~\cite{Olmo:2005hc} that laboratory tests on the macroscopic Euler equation could put strong limits on $a_2$.

One can also find a conserved momentum quantity in Palatini gravity. To this end, first one needs to consider the 1PN corrections to the Euler equation. The computational details can be found in Appendix \ref{sec:math}.
At the end, one finds the following vector conservation law,
\begin{equation}
\frac{d}{dt}P^{j}=0,
\end{equation}
with
\begin{align}
P^{j}=\int\rs v^{j}\bigg(1 + \frac{v^{2}}{2} - \frac{U}{2} + \Pi &+ \frac{3p}{\rs}-\aii\rs\bigg)d^{3}x \ - \notag \\[1ex]
&- \frac{1}{2}\int\rs W^{j}d^{3}x. \label{momentum}
\end{align}
This is the total momentum which is conserved in Palatini gravity. This conservation law depends on the PN Euler equation, in the latter it is not possible to replace Palatini contributions with effective expressions for $p$ and $\Pi$. If it was possible, all the Palatini effects could be interpreted as redefinitions of perfect fluid quantities.

The previous results show that Palatini $f(R)$ gravity does not violate total conservation of energy and momentum in the PN regime. This is an expected outcome since any Palatini gravity model is a Lagrangian based metric theory with matter action being independent from the affine connection and, as shown in \cite{Lee:1974nq}, they should not violate PN conservation laws. In the context of the PPN formalism, the results obtained here directly show that the PPN parameters $\zeta_1,\zeta_2,\zeta_3,\zeta_4$ and $\alpha_3$ are all zero in Palatini $f(R)$ gravity. The expressions found here show explicitly what are the conserved quantities.

\section{Equation of motion for massive bodies} \label{sec:eqofmot}
In this section we split the fluid description of the source into $N$ separated bodies. We aim to obtain the PN equations of motion for the bodies center-of-mass positions. This is a realistic way to deal with the trajectories of massive and finite volume bodies, instead of assuming test particles. Each body indexed by $A$ has a material mass given by
\begin{equation}\label{massa}
m_A=\int_{A}\rs d^{3}x.
\end{equation}
The volume where the integration above is calculated is a time-independent region of space that extends beyond the volume occupied by the body. It is large enough that, in a time interval $dt$, the body does not cross its boundary surface but it is also small enough to not intersect with any other body of the system. The center-of-mass of a body $A$, its velocity and acceleration is then defined as
\begin{align}
	\bs{r}_A(t)\equiv& \ \dfrac{1}{m_A}\int\rs\bs{x}\,d^{3}x,\\[2ex]
	\bs{v}_A(t)\equiv& \ \dfrac{d\bs{r}_A}{dt}=\dfrac{1}{m_A}\int\rs\bs{v}\,d^{3}x,\\[2ex]
	\bs{a}_A(t)\equiv& \ \dfrac{d\bs{v}_A}{dt}=\dfrac{1}{m_A}\int\rs\dfrac{d\bs{v}}{dt}\,d^{3}x \,.  \label{a} 
\end{align}
In the above, we used the property \eqref{dt}. Following \cite{PoissonWill}, the center-of-mass acceleration of each body is decomposed into three parts,
\begin{equation}\label{acc}
\bs a_A= \bs a_{A}^{\mbox{\tiny Newt}} +\bs a^{\mbox{\tiny PN}}_A + \bs a^{\mbox{\tiny Str}}_A \, .
\end{equation}
The first is the Newtonian contribution, the second is the PN contribution apart from structural  contributions. Considering the solar system application, $\bs a^{\mbox{\tiny Str}}_A$ only depends on interactions among the planets and the planet into itself. The interaction of a planet into itself is  relevant for studying the planet stability (which is not our purpose here) and it may also be relevant for the center-of-mass acceleration. The latter is not the case of GR, but it is relevant for some gravitational theories (e.g.,  the Nordtvedt effect \cite{Nordtvedt:1968qs, Will:1971zzb}).

The integrand of eq.~\eqref{a} is found from the Euler equation with its PN extension, which comes from eq.~\eqref{divt}, and it reads
\begin{align}\label{pneuller}
\rs\dfrac{dv^{j}}{dt}=& \ \, \rs\p^{j}U-\p^{j}p +\left(\dfrac{v^{2}}{2}+U+\Pi+\dfrac{p}{\rs}\right)\p_{j}p \ -\notag\\
& - v^{j}\p_tp+\rs\bigg[(v^{2}-4U)\p_{j}U +4\p_tV^{j} +\p_j\psi \ + \notag \\
& +4v^{k}\p_{[k}V_{j]} +\dfrac{1}{2}\p_{ttj}\chi -v^{j}(3\p_tU+4v^{k}\p_kU) \bigg] \,-\notag\\
& -\aii\rs\left(\dfrac{v^{2}}{2}+U+\Pi\right)\p_j\rs +\dfrac{\aii}{2}\,v^{j}\p_t(\rs^{2})\,+ \notag\\
& +\aii\p_j\left[\rs{}^{2}\left(\dfrac{1}{2}- \dfrac{v^{2}}{2}+3U-\Pi\right)\right] \, -\notag\\[1ex]
& -(2\aii^{2}+\aiii)\p_j(\rs{}^{3}) -2\aii\rs\p_j\Phi_P .
\end{align}

We use  eq.~\eqref{pneuller} in eq.~\eqref{a} and compute the integrals to obtain the center-of-mass acceleration. The integration techniques are detailed and discussed in Appendix \ref{sec:math}. The final results for the accelerations $\bs a_{A}^{\mbox{\tiny Newt}}, \bs a^{\mbox{\tiny PN}}_A$ and $\bs a^{\mbox{\tiny Str}}_A$ read 

\begin{equation}
\bs a_{A}^{\mbox{\tiny Newt}} = -\sum_{B\neq A}\dfrac{m_B}{r_{AB}^{2}}\,\bs n_{AB}\, , \label{anewt}
\end{equation}
\begin{eqnarray}
	\bs a_{A}^{\mbox{\tiny PN}}&=& -\sum_{B\neq A}\dfrac{m_B}{r_{AB}^{2}}\bigg\{
	 \bigg[ v_A^{2}-4(\bs v_A\cdot\bs v_B)+2v_B^{2}\,-\notag\\[1ex]
	 && - \ \dfrac{3}{2}(\bs n_{AB}\cdot\bs v_B)^{2} -\dfrac{5m_A}{r_{AB}} -\dfrac{4m_B}{r_{AB}} \bigg] \bs n_{AB} \,-\notag\\[1ex]
	 && - \ \big[ \bs n_{AB}\cdot (4\bs v_A-3\bs v_B) \big] (\bs v_A -\bs v_B)\,+ \notag\\[1ex]
	&& + \ \dfrac{7}{2}\sum_{C\neq A,B}\!\! m_C\dfrac{r_{AB}}{r_{BC}^{2}}\,\bs n_{BC} \ +  \label{apn} \\[1ex]
	 && \, - \sum_{C\neq A,B}m_C\left[ \dfrac{4}{r_{AC}} +\dfrac{1}{r_{BC}} \right. \ -\notag\\[1ex]
	 && \,\left.  - \ \dfrac{r_{AB}}{2r_{BC}^{2}}(\bs n_{AB}\cdot\bs n_{BC}) \right]\bs n_{AB} \nonumber
	  \bigg\} \, ,
\end{eqnarray}
\begin{equation}
	\bs a_{A}^{\mbox{\tiny Str}} = -\sum_{B\neq A}\dfrac{E_{B}}{r_{AB}^{2}}\,\bs n_{AB} \, . \label{str}	
\end{equation}

In the above expressions, we use the definitions $\bs r_{AB}=\bs r_A -\bs r_B$, $r_{AB}=|\bs r_{AB}|$ and $\bs n_{AB}=\bs r_{AB}/r_{AB}$. These equations have no explicit dependence on either $a_2$ or $a_3$ and they are identical, in form, to the corresponding GR expressions \cite{PoissonWill}. There is a single implicit difference, it is inside the constant $E_B$, which is the energy associated with the planet indexed with $B$. Indeed, according to eq.~\eqref{energy}, the energy expression has a correction that depends on $a_2$. However, using the effective internal energy $\Pi_{\mbox{\tiny eff}}$ \eqref{PiEff}, the $a_2$ dependence also vanishes and the center-of-mass acceleration expressions become {\it identical} to the GR ones. Similar results, in the context of compact objects and using other methods, were  obtained in Ref.~\cite{Enqvist:2013tsa}, where the orbital motion of binary systems was found to be the same as in GR, apart from a mass redefinition.

Although the metric in Palatini $f(R)$ gravity depends on terms that are outside the standard PPN formalism, the above equations of motion can be directly compared to the  general equations of motion of the formalism (Appendix \ref{ppn}). Therefore, considering the physical interpretation of the PPN parameters, one concludes that  the remaining PPN parameters are $\beta=1$ and $\alpha_1=\alpha_2=\xi=0$. Hence the values of all the PPN parameters are the same as GR. To be clear, this does not imply that Palatini $f(R)$ theories, with analytical $f(R)$, are identical to GR, only that the set of tests performed in the PPN context (which include solar system tests) is not sensitive to such differences.

\section{Discussion on a heuristic argument for the orbits of test particles and massive bodies} \label{sec:discussions}

In Ref.~\cite{Will:1971zzb} there is an argument, suggested by Richard Price, on why the internal structure of planets does not influence their center-of-mass orbits in GR, while their internal structures can influence other theories, like scalar-tensor and metric $f(R)$ theories through the Nordtvedt effect. We comment here that the same argument also holds for Palatini $f(R)$ gravity whenever the  $f(R)$ function is such that it admits solutions for the planets that are sufficiently close to being asymptotically  flat. This both presents a more intuitive understanding for our results and further extends them towards some non-analytical $f(R)$ cases. The argument goes as follows \cite{Will:1971zzb}: {\it ``Consider a massive body located in an external gravitational field which can be considered uniform over a region that is very large compared with the body’s gravitational radius. Focus attention on a large volume of space V surrounding the massive body --- a volume so large that in its outer regions the spacetime curvature produced by the body itself is small to some desired accuracy (asymptotic flatness); but a volume still small enough that throughout it the external gravitational field is homogeneous to some other desired accuracy. In the outer regions of V one can introduce an inertial reference frame that falls freely in the homogeneous external field. Of course, that inertial frame cannot be extended into the massive body; but it does completely surround the body (asymptotic flatness). Conservation of the body’s total four-momentum in that frame (valid for any massive body in asymptotically flat spacetime) guarantees that, if it is initially at rest in our inertial frame, it will always remain at rest there. Similarly, any test particle (far from the massive body) initially at rest will remain at rest. Thus both test particles and the massive body are tied to the inertial frame. This means that, as seen in the original accelerated frame where the external field is manifest, they fall with identical accelerations.''}

Although GR was the original context of the argument above, it can be applied to a large class of Palatini $f(R)$ theories. This class needs not be analytical but needs to admit solutions that are sufficiently close to being asymptotically flat, as described above. When these conditions are met, and since there is energy-momentum conservation in Palatini gravity (Section \ref{sec:conserved}), test particles in Palatini $f(R)$ will describe the same center-of-mass orbits of the massive-finite-volume planets. Also, since test particles have the same geodesics of GR, considering the internal energy redefinition as detailed in Section \ref{palatini}, the PPN parameters of this Palatini $f(R)$ class of theories will have the GR values. 

Considering the argument above, unless there is a physical procedure capable of testing the internal energy redefinition \eqref{PiEff},  or unless the approximation on asymptotically flatness fails, GR and Palatini $f(R)$ yield exactly the same physics for the solar system orbits. For the first case, evaluating  gravitational effects from the internal energy at atomic level may be a fruitful route, and there is a current debate on this subject within GR (e.g., \cite{Zych:2011hu, Pikovski:2013qwa, Schwartz:2019qqg}).  {\it A priori}, and assuming the existence of some interpretation in which it would be possible to physically distinguish $\Pi$ and $\Pi_\eff$, if future experimental results imply that the internal energy of the system does not generate the expected amount of gravitational effects, then the Palatini $f(R)$ energy shift \eqref{PiEff} may be used to explain the mismatch. In order to study microscopic properties of matter with Palatini gravity, it may be relevant to couple Palatini gravity to the Dirac equation. This type of coupling is not covered by our analysis, since it corresponds to a $S_{\mbox{\tiny matter}}$ that depends on the connection $\Gamma$. According to Ref.~\cite{Olmo:2008ye}, which considers the Dirac equation and  $f(R) = R + \alpha_2 R^2$, the parameter $\alpha_2$ can be constrained in this setting.

\section{Conclusion}\label{sec:conclusion}

Here we present  a detailed post-Newtonian analysis of Palatini $f(R)$ dynamics at the solar system, and we considered the restriction to analytic $f(R)$ expressions. The latter restriction is essential for finding a picture closer to the standard Will-Nordtvedt PPN formalism \cite{will1993theory, Will:2014kxa}. Considering the previous analyses of Palatini $f(R)$ gravity in the solar system context \cite{Allemandi:2005tg,Sotiriou:2005xe,Olmo:2005hc,Olmo:2005zr}, our work does not use the equivalence between $f(R)$ and scalar-tensor theories and, more importantly, we test directly the equations of motion, which is a necessary step since Palatini $f(R)$ gravity leads to new terms in the metric that are not part of  the 10 potentials from the standard Will-Nordtvedt PPN formalism. 

We find that the equations of motion of the center-of-mass orbits are precisely the same as in  GR, even in the case that  the planets are seen as massive bodies of finite volume. To achieve this conclusion, redefinitions of energy and momentum were necessary [see (\ref{PipEff}), (\ref{PiEff}) and (\ref{pEff})]. These are not sensitive to the type of experiment performed for the solar system orbits, but in principle may be constrained by atomic level experiments, as discussed in Section \ref{sec:discussions}. 

For analytic $f(R)$ functions with negligible $\Lambda_\eff$ \eqref{Leff}, and in contrast to some results of Ref.~\cite{Olmo:2005hc}, we find no restriction on the coefficients of $R^2$ and $R^3$ based on either the orbits of the planets  or on (macroscopic) laboratory tests of the Euler equation. The latter since we find no reason to interpret eq.~\eqref{neuller} as a modified Euler equation, we understand it as the Euler equation itself, but written in a different form. Hence, from the tests here considered, we find no need for demanding that $f(R)$, within analytical Palatini gravity, should be close to a linear function: the coefficients of $R^2$ and higher powers on $R$ can be arbitrary. As commented in Sec.~\ref{sec:discussions}, these results can be extended towards some non-analytical $f(R)$ functions, this whenever the spacetime can be considered, within some approximation, as being asymptotically flat far from the matter distributions. 

Our conclusions, as above explained, do not apply in general to $f(R)$ functions that include the term $1/R$; unless the corresponding coefficient of the $1/R$ term is sufficiently small, in order to assure an asymptotic-flatness-like behavior up to some radius. We remark that some works in the literature (e.g., \cite{Olmo:2005hc, Olmo:2006zu}) show that the $1/R$ term can introduce considerable differences with respect to GR, implying that there are physical constraints on the value of the $1/R$ coefficient. Hence, with respect to the $1/R$ term, there is no incompatibility with our results.

\acknowledgements  We thank Gonzalo Olmo for useful comments on a draft version of this work. A.W. thanks the Department of Theoretical Physics and IPARCOS at Complutense University of Madrid for its hospitality during different stages of the elaboration of this work. J.D.T. and A.W. thank FAPES and CAPES (Brazil) for support. D.C.R. thanks CNPq and FAPES (Brazil) for partial support.

\appendix

\section{On the derivation of the conserved momentum and the center-of-mass acceleration} \label{sec:math}

In this appendix we detail the main mathematical procedures used in Sections \ref{sec:conserved} and \ref{sec:eqofmot}. We start by the procedure to obtain the conserved momentum present in eq.~\eqref{momentum}. Taking the PN contribution in eq.~\eqref{divt}, one can write,
\begin{align}
0=& \ \p_t(\mu\rs v^{j})+\p_{k}(\mu\rs v^{k}v^{j}) + 2\p_{j}(Up) \ -\notag\\[1ex]
&  - \,\rs\left(\frac{3}{2}\,v^{2}-U+\Pi+ 3\frac{p}{\rs} \right)\p_{j}U + 4\rs v^{k}\p_{j}V_{k}\, +\notag\\[1ex]
& + 2\rs\frac{d}{dt}\left(Uv^{j} -2V^{j}\right)- \rs\p_{j}\psi -\dfrac{1}{2}\rs\p_{ttj}\chi \ +\notag\\[1ex]
& +\aii\bigg[-2\rs\frac{d}{dt}(\rs v^{j}) +2\rs{}^{2}\p_{j}U \  + 2\rs\p_{j}\Phi_P \ +\notag\\[1ex]
& + \p_{j}\left(2\rs{}^{2}U-\rs p +\frac{1}{2}\rs{}^{2}v^{2}\right. \ -\notag\\[1ex]
& \left. - \, \rs{}^{2}\Pi +\frac{5}{3}\aii \rs^{3}+\frac{\aiii}{\aii}\rs{}^{3} \right) \bigg]. \label{1PNEuler}
\end{align}
In the above, the Euler equation \eqref{neuller} was used and
\begin{equation}
\mu\equiv \frac{v^{2}}{2} + U +\Pi+ \frac{p}{\rs} +\aii\rs.
\end{equation}
We now integrate over the volume occupied by the fluid. All the terms not proportional to $\aii$ are standard GR terms and their integration are detailed in \cite{PoissonWill}. The first term inside the square bracket is solved through relation \eqref{dt}, while the second and third ones cancel each other when integrated. The latter can be verified using the ``switch trick'', which consists in interchanging the variables $x\leftrightarrow x'$ inside the integral,
\begin{align}
\int\rs\p_{j}\Phi_Pd^{3}x=& \int\rs\rs{}'{}^{2}\p_j(|\bs x- \bs x'|^{-1})d^{3}x'd^{3}x  \notag\\[1ex]
 \overset{\leftrightarrow}{=}&\int\rs{}'\rs{}^{2}\p'_j(|\bs x'- \bs x|^{-1})d^{3}xd^{3}x' \, -\notag\\[1ex]
  &-\int\rs{}^{2}\rs{}'\p_j(|\bs x- \bs x'|^{-1})d^{3}x'd^{3}x\notag\\[1ex]
  =& -\int\rs{}^{2}\p_{j}Ud^{3}x, \label{idx}
\end{align}
where the symbol $\overset{\leftrightarrow}{=}$ indicates when the switch trick was performed and the identity $\p_j[f(x-x')]=-\p'_j[f(x-x')]$ was also used.

With the above results, one easily finds the Palatini contribution in eq.~\eqref{momentum}.

Let us now analyze the integration of eq.~\eqref{pneuller}, in order to obtain the PN expression for the body's acceleration. First, we use Euler's equation \eqref{neuller} to express the second time derivative of $\chi$ in terms of other potentials, namely
\begin{equation}\label{ttchi}
\p_{tt}\chi= \Phi_{1}+2\Phi_4- \Phi_{5} -\Phi_6 -\aii \Phi_P,
\end{equation}
where,
\begin{align}
	\Phi_{5}=& \ \int\rs{}'\p'_{k}U'\,\dfrac{(x-x')^{k}}{\vert\bs{x}-\bs{x}'\vert}\,d^{3}x',\\[2ex]
	\Phi_{6}=& \ \int\rs{}' v'_{k}v'_{j}\,\dfrac{(x-x')^{k}(x-x')^{j}}{\vert\bs{x}-\bs{x}'\vert^{3}}\,d^{3}x'.
\end{align}
To integrate eq.~\eqref{pneuller}, it is introduced relative variables with respect to the center-of-mass, said $\bs\bx=\bs x- \bs r_A$ and $\bs\bv=\bs v- \bs v_A$. Each body of the system is assumed to be reflection-symmetric above its own center-of-mass, for example $\rs(t,\bs r_A-\bs\bx)=\rs(t,\bs r_A+\bs\bx)$, and the same for the pressure $p$ and the internal energy density $\Pi$. This symmetry allows to eliminate any integral having an odd number of internal vectors, such as $\bs\bx$, $\bs\bv$ or $\bs\nabla p$. Also, the gravitational potentials are separated into an internal part, produced by the body $A$, and an external one sourced by the remaining bodies of the system. For instance,
\begin{align}
	\Phi_P&=\int_{A}\dfrac{\rs{}'^{2}}{|\bs x-\bs x'|}\,d^{3}x' + \sum_{B\neq A}\int_B\dfrac{\rs{}'^{2}}{|\bs x-\bs x'|}\,d^{3}x',\notag\\[2ex]
	&\equiv \Phi_{P,A} + \Phi_{P,A}^{ext}.\label{ext}
\end{align}
When integrating the parts containing the internal pieces both integrals will have the same domain, so they can be computed with the help of the switch trick mentioned before. Assuming a wide separation between bodies implies that, when evaluating an external potential within the body $A$, it can be expanded in a Taylor series. As an example, one has
\begin{equation}
\Phi_{P,A}^{ext}(t,\bs x)\a \Phi_{P,A}^{ext}(t,\bs r_A)+ \bx^{j}\p_j\Phi_{P,A}^{ext}(t,\bs r_A)  +\dots\,.
\end{equation}
The series is truncated in its second term since the next terms are at least of order $(R_A/r_{AB})^{2}$, where $R_A$ is the typical body radius and $r_{AB}=|\bs r_A -\bs r_B|$ is the interbody distance. This expansion is used to extract the external pieces of potentials from the integrals. At the end, the acceleration of the center-of-mass of a body $A$ can be written as
\begin{align}
	a_A^{j}=& \ \p_jU_A^{ext} - \dfrac{1}{m_A}\bigg[\left(4H_A^{(jk)}-3K_A^{jk}+\dot{{\cal P}}_A\delta^{jk} \right. \ +\notag\\[1ex]
+& \left.\dot{E}_{P,A}\delta^{jk}-2L_A^{(jk)}-2L_{P,A}^{(jk)}\right)v_A^{k} \ +\notag \\[1ex]
	 +& \,4\left(2{\cal T}_A^{jk}+ \Omega_{A}^{jk}+ {\cal P}_A\delta^{jk}+E_{P,A}\delta^{jk}\right)\p_k U^{ext}_A \ +\notag\\[1ex]
	 -& \, \left(2{\cal T}_A+ \Omega_{A}+ 3{\cal P}_A+3E_{P,A}\right)\p_j U^{ext}_A \bigg] \ + \notag \\[1ex]
	 +& \, \dfrac{1}{c^{2}}\bigg[(v_A^{2}-4U_A^{ext})\p_jU_A^{ext}- v_A^{j}(4v_A^{k}+3\p_tU_A^{ext}) \, - \notag\\[1ex]
	-& \, 4v_A^{k}\p_{[j}V_{k],A}^{ext} +4\p_tV_{j,A}^{ext} +\p_{j}\psi_A^{ext} -\dfrac{5\aii}{2}\,\p_j\Phi_{P,A}^{ext} \bigg],\label{aext}
\end{align}
where
\begin{align}
        {\cal T}_{A}=& \ \frac{1}{2}\int_{A}\rs \bv^{2}d^{3}\bx,\\[1ex]
        \Omega_{A}=& - \frac{1}{2}\int_{A}\rs Ud^{3}\bx,\\[1ex]
        E_{P,A}=& -\frac{\aii}{2}\int_{A}\rs{}^{2}d^{3}\bx,\\[1ex]
	{\cal P}_A=&\int_A pd^{3}\bx,\\[1ex]
	{\cal T}_A^{jk}=& \ \dfrac{1}{2}\int_A\rs\bv^{j}\bv^{k}d^{3}\bx,\\[1ex]
	\Omega_A^{jk}=&-\dfrac{1}{2}\int_A\rs\rs{}'\dfrac{(\bx-\bx')^{j}(\bx-\bx')^{k}}{|\bs\bx-\bs\bx'|^{3}}\,d^{3}\bx'd^{3}\bx,\\[1ex]
	H_A^{jk}=& \int_A\rs \rs{}'\dfrac{\bv'^{j}(\bx-\bx')^{k}}{|\bs\bx-\bs\bx'|^{3}}\,d^{3}\bx' d^{3}\bx,\\[1ex]
	K_A^{jk}=& \int_A\rs\rs{}'\dfrac{\bv'_n(\bx-\bx')^{n}(\bx-\bx')^{j}(\bx-\bx')^{k}}{|\bs\bx-\bs\bx'|^{5}}\,d^{3}\bx'd^{3}\bx,\\
	L_A^{jk}=& \int_{A}\bv^{j}\p^{k}p \,d^{3}\bx,\\[1ex]
	L_{P,A}^{jk}=&-\dfrac{\aii}{2}\int_{A}\bv^{j}\p^{k}(\rs{}^{2}) d^{3}\bx.
\end{align}
are called structure integrals and the external potentials must be evaluated at $\bs x=\bs r_A$, after differentiation.

With the assumption that any modification in the internal structure  of each body occurs only in a timescale much longer than its orbital period, it is possible to derive equilibrium conditions from the time independence of the quadrupole moment of mass distribution \cite{PoissonWill}. The relevant expressions here are the following ones,
\begin{align}
&2{\cal T}^{jk}_{A}+{\cal P}_{A}\delta^{jk}+\Omega^{jk}_{A}+E_{P,A}\delta^{jk}=0,\label{eq1}\\[2ex]
&2{\cal T}_{A}+3{\cal P}_{A}+\Omega_{A}+3E_{P,A}=0,\label{eq2}\\[2ex]
&4H^{j(k)}_{A}-3K_{A}^{(jk)}+\dot{{\cal P}}_{A}\delta^{jk}+\dot{E}_{P,A}\delta^{jk} \ -\notag\\[1ex]
&\quad -2L_{A}^{(jk)}-2L_{P,A}^{(jk)}=0,\label{eq3}
\end{align}
where it was also assumed that the bodies are not spinning. Expressions above then eliminate the self-interaction terms from $\bs a_A$.

The next task is to express the external potentials explicitly in terms of the center-of-mass positions and their velocities.
Exemplifying with $\p_j\Phi_{P,A}^{ext}$, one has
\begin{align}
\dfrac{\aii}{2}\p_j \Phi_{P,A}^{ext}&= -\dfrac{\aii}{2}\sum_{B\neq A}\int_B\dfrac{\rs{}'^{2}(r_A-x')^{j}}{|\bs r_A -\bs x'|^{3}}d^{3}x' \notag\\[1ex]
&=-\dfrac{\aii}{2}\sum_{B\neq A}\int_B\dfrac{\rs{}'^{2}(r_{AB}-\bx')^{j}}{|\bs r_{AB} -\bs \bx'|^{3}}d^{3}\bx',
\end{align}
It is used, once more, that the bodies are widely separated and the terms $|\bs r_{AB}-\bs \bx'|$ are expanded in powers of $\bx'$, yielding
\begin{align}
\dfrac{\aii}{2}\p_j \Phi_{P,A}^{ext}=& -\dfrac{\aii}{2}\sum_{B\neq A}\dfrac{n_{AB}^{j}}{r_{AB}^{2}}\int_B\rs{}'^{2}d^{3}\bx' \notag \\[1ex]
=& \sum_{B\neq A}\dfrac{E_{P,B}\,n_{AB}^{j}}{r_{AB}^{2}},
\end{align}
where $\bs n_{AB}=\bs r_{AB}/r_{AB}$. Only two terms in the series expansion were considered, since the other ones are at least of order $(R_A/r_{AB})^{2}$ and can be neglected. Also, the second term leads to a vanishing integral due to an odd number of internal vectors.

Proceeding in similar lines with the others potentials, the final expressions given in \eqref{anewt}-\eqref{str} are then achieved.

\section{Metric and equations of motion in the PPN formalism}\label{ppn}

The modern PPN formalism, as presented by Will and Nordtvedt, proposes a general metric expanded in terms of gravitational potentials and ten constant parameters,
\begin{align}
	g_{00} = -&1 + 2U - 2\beta U^2 + (2 \gamma +2+\alpha_3 +\zeta _1-2 \xi ) \Phi_1 \,+\notag\\[1ex]
	+& \, 2(3 \gamma -2\beta+1+\zeta _2+ \xi ) \Phi_2 \ + 2(1+\zeta _3 ) \Phi_3 \,+ \notag\\[1ex]
      +& \, 2(3 \gamma +3\zeta _4-2 \xi ) \Phi_4 - (\zeta _1-2 \xi )\Phi_6 \, - \notag\\[1ex]
      -& \, 2\xi \Phi_W + (1+\alpha_2 -\zeta_1+2\xi)\p_{tt}\chi \nonumber \\[2ex]
	g_{0i} = -&(2\gamma +2+\alpha_1/2) V_i \label{34} \\[2ex]
	g_{ij} = \ (&1+2\gamma \,U)\, \delta_{ij}\,.\nonumber
\end{align}

The above metric is in the gauge used in ref.~\cite{PoissonWill}, the same worked in this paper. The particular way in which the PPN coefficients
are presented ensures specific physical meanings to each of them (see Ref~\cite{will1993theory})

Using the PN metric the equations of motion are derived. In the case of photons, it is used the geodesic equation, yielding
\begin{equation}\label{eqlight}
\dfrac{dn^{j}}{dt}=(1+\gamma)\left(\delta^{jk}-n^{j}n^{k}\right)\p_kU,
\end{equation}
where $n^{j}$ is a unitary vector for the velocity of the photon. For massive bodies, the procedure is the same developed before and the acceleration of a body $A$ can be written as $\bs a_A=\bs a_{A}^{\mbox{\tiny Ext}} + \bs a_A^{\mbox{\tiny Self}}$. The $\bs a_{A}^{\mbox{\tiny Ext}}$ contains only the effects produced by the other bodies of the system. Thus, it includes the results found in eqs.~\eqref{anewt}-\eqref{str} for Palatini gravity. The $\bs a_A^{\mbox{\tiny Self}}$ part accounts for the self-acceleration term since it depends on the internal structure of the body $A$. Their respective expressions are shown below.
\begin{widetext}
\begin{align}
	\bs a_{A}^{\mbox{\tiny Ext}}=& -\sum_{B\neq A}\dfrac{M_B}{r_{AB}^{2}}\bigg\{
	\bigg[1+\gamma v_A^{2}-(2\gamma+2+\tfrac{1}{2}\alpha_1)(\bs v_A\cdot\bs v_B) +(\gamma+1+\tfrac{1}{2}\alpha_2+\tfrac{1}{2}\alpha_3)v_B^{2} \ - \notag\\
	&-\dfrac{3}{2}(1+\alpha_2)(\bs n_{AB}\cdot\bs v_B)^{2} - (2\gamma +2\beta+1+\tfrac{1}{2}\alpha_1-\zeta_2)\dfrac{m_A}{r_{AB}} -(2\gamma+2\beta)\dfrac{m_B}{r_{AB}} \bigg] \bs n_{AB} \,-\notag\\
	&-	\bs n_{AB}\cdot [(2\gamma+2)\bs v_A-(2\gamma+1)\bs v_B]\bs v_A \ + \notag\\
	& +\bs n_{AB}\cdot [(2\gamma+2+\tfrac{1}{2}\alpha_1)\bs v_A -(2\gamma+1+\tfrac{1}{2}\alpha_1-\alpha_2)\bs v_B]\bs v_b \ +\notag\\
	&+\dfrac{1}{2}(4\gamma+3-2\xi+\alpha_1-\alpha_2+\zeta_1)\sum_{C\neq A,B}\!\! M_C\dfrac{r_{AB}}{r_{BC}^{2}}\,\bs n_{BC} \ -\\
	&- \sum_{C\neq A,B}M_C\bigg[(2\gamma+2\beta-2\xi) \dfrac{1}{r_{AC}} +(2\beta-1-2\xi-\zeta_2)\dfrac{1}{r_{BC}} \ -\notag\\
	&-\dfrac{1}{2}(1+2\xi+\alpha_2-\zeta_1)\dfrac{r_{AB}}{r_{BC}^{2}}(\bs n_{AB}\cdot\bs n_{BC}) -\xi\dfrac{r_{BC}}{r_{AC}^{2}}(\bs n_{AC}\cdot\bs n_{BC})\bigg]\bs n_{AB} \ -\notag\\
	& -\xi\sum_{C\neq A,B}\dfrac{M_C}{r_{AB}}\left[(\bs n_{AC}-\bs n_{BC})-3\bs n_{AB}\cdot(\bs n_{AC}-\bs n_{BC})\bs n_{AB}\right]	\bigg\},\notag
\end{align}

\begin{align}
	a_A^{\mbox{\tiny Self},\,j}=& \sum_{B\neq A}M_B\bigg[(4\beta-\gamma-3-4\xi-\alpha_1 +\alpha_2-\zeta_1)\dfrac{\Omega_A}{M_A}\p_j\left(\dfrac{1}{r_{AB}}\right) \ -\notag\\
	&-(\alpha_2-\zeta_1+\zeta_2)\dfrac{\Omega_A^{jk}}{M_A}\p_k\left(\dfrac{1}{r_{AB}}\right) +\xi\dfrac{\Omega_A^{kl}}{M_A}\p_{jkl}(r_{AB}) \ +\\
	&+(4\beta-\gamma-3-4\xi-\tfrac{1}{2}\alpha_3 -2\zeta_2)\dfrac{\Omega_B}{M_B}\p_j\left(\dfrac{1}{r_{AB}}\right) + (\xi-\tfrac{1}{2}\zeta_1)\dfrac{\Omega_B^{kl}}{M_A}\p_{jkl}(r_{AB}) \ -\notag\\
	&-\zeta_3\dfrac{E_B^{int}}{M_B}\p_j\left(\dfrac{1}{r_{AB}}\right) + (\tfrac{3}{2}\alpha_3-3\zeta_4+\zeta_1)\dfrac{P_B}{M_B}\p_j\left(\dfrac{1}{r_{AB}}\right)\bigg] -\dfrac{\alpha_3}{M_A}H_A^{jk}v_{A,k}.\notag
\end{align}
\end{widetext}

%\bibliography{AllMyRefs.bib}


\end{document}